%% file: main.tex
\documentclass[acmsmall,preprnt]{acmart}

\AtBeginDocument{%
  }

\input{commands}
\input{figures}

\begin{document}

\title{Fostering Enterprise Conversations Around Data on Collaboration Platforms}

\author{Hyeok Kim}
\email{hyeok@northwestern.edu}
\affiliation{%
  \institution{Northwestern University}
  \city{Evanston}
  \state{Illinois}
  \country{USA}
}

\author{Arjun Srinivasan}
\email{arjun.srinivasan@databricks.com}
\affiliation{%
  \institution{Tableau Research}
  \city{Seattle}
  \state{Washington}
  \country{USA}
}

\author{Matthew Brehmer}
\email{mbrehmer@uwaterloo.ca}
\affiliation{%
  \institution{University of Waterloo}
  \city{Waterloo}
  \state{Ontario}
  \country{Canada}
}

\renewcommand{\shortauthors}{Kim et al.}

\input{sections/0-abstract}

\maketitle

\input{sections/1-introduction}
\input{sections/2-background}
\input{sections/3-co-design}
\input{sections/4-scenario}
\input{sections/5-interview}
\input{sections/6-discussion}

\begin{acks}
The authors conducted this work while affiliated with Tableau. A citable version of this work is available via \url{https://doi.org/10.1109/VIS55277.2024.00024}.
\end{acks}

\bibliographystyle{ACM-Reference-Format}
\bibliography{reference}


\end{document}

%% file: commands.tex
\usepackage{xargs} % Used for new commands with optional arguments
\usepackage{soul}  % Used for custom comments
\usepackage{color} % Used for custom colors in comments
\usepackage{xspace}
\usepackage{listings}

\usepackage[vflt]{floatflt}

\newcommand{\ie}{{i.e.,}\xspace}
\newcommand{\eg}{{e.g.,}\xspace}

\newcommand{\ea}{{et~al.}\xspace}

\newcommand{\bpstart}[1]{\vspace{1mm} \noindent{\textbf{#1.}}}
\newcommand{\ipstart}[1]{\vspace{1mm} \noindent{\textit{#1.}}}

\definecolor{activegold}{RGB}{255,193,61}

\definecolor{lightorange}{rgb}{1,0.8,0.4}
\definecolor{lightorange}{RGB}{230, 170, 50}
\definecolor{lightgreen}{RGB}{121,210,121}
\definecolor{lightteal}{RGB}{121,199,210}
\definecolor{lightblue}{RGB}{100,212,239}
\definecolor{lightpurple}{RGB}{153,102,255}
\definecolor{lightred}{RGB}{245, 132, 120}
\definecolor{red}{RGB}{178,34,34}
\definecolor{gray}{RGB}{166,166,166}
\definecolor{indexBlue}{cmyk}{0.9,0.8,0,0}
\definecolor{indexGreen}{cmyk}{0.8,0.2,0.8,0.55}
\definecolor{lightgray}{RGB}{185,185,185}
\definecolor{deepblue}{cmyk}{0.9,0.75,0,0.5}
\definecolor{deepred}{cmyk}{0,0.75,0.75,0.4}
\definecolor{pink}{RGB}{214,114,0}

\newcommandx{\todo}[1][]{
    \ifnotes
        {\textcolor{red}{[TODO] \emph{#1}}}
    \else
    \fi
}

\newcommandx{\cut}[1][]{
    \if\reviewmode
        {\textcolor{red}{\st{#1}}}
    \else
    \fi
}

\newcommandx{\sout}[1][]{
    \ifnotes
        {\textcolor{red}{\st{#1}}}
    \else
    \fi
}

\newcommandx{\claim}[1][] {\ifnotes {\textbf{\textcolor{pink}{Claim:} #1}}\else#1\fi}

\newcommandx{\guest}[3][1=]{
    \ifnotes
        \setulcolor{lightorange}{\ul{#1}} \textcolor{lightorange}
        {[\textbf{#2:} #3]}
    \else
        #1
    \fi
}
\newcommandx{\matt}[2][1=]{
    \ifnotes
        \setulcolor{lightred}{\ul{#1}} \textcolor{lightred}
        {[\textbf{Matt:} #2]}
    \else
        #1
    \fi
}

\newcommandx{\arjun}[2][1=]{
    \ifnotes
        \setulcolor{lightgreen}{\ul{#1}} \textcolor{lightgreen}
        {[\textbf{Arjun:} #2]}
    \else
        #1
    \fi
}

\newcommandx{\hyeok}[2][1=]{
    \ifnotes
        \setulcolor{lightblue}{\ul{#1}} \textcolor{lightblue}
        {[\textbf{Hyeok:} #2]}
    \else
        #1
    \fi
}

\newcommandx{\revised}[1][]{
    \ifnotes
        {\textcolor{blue}{#1}}
    \else
        #1
    \fi
}
\definecolor{snapBlue}{cmyk}{0.67,1,0,0.18}
\definecolor{compGreen}{cmyk}{0.92,0,1,0.18}
\definecolor{tempOrange}{cmyk}{0,0.27,0.97,0.24}
\definecolor{tempRed}{cmyk}{0,0.69,0.91,0.32}

\newcommandx{\revision}[2][1=]{\iftrack{\textcolor{lightgray}{\st{#1}}\textcolor{red}{#2}}\else{#2}\fi}
\newcommandx{\removed}[1][]{\iftrack{\textcolor{red}{\st{#1}}}\else{}\fi}

\usepackage{tcolorbox}
\definecolor{sceneNoCol}{cmyk}{0,0,0,0.4}
\newtcbox{\sceneNo}{nobeforeafter,tcbox raise base, %inline
boxrule=0mm,top=0mm,bottom=0mm,right=0.25mm,left=0.25mm,arc=1.5mm,boxsep=1mm,
fontupper=\sffamily\small,
colframe=sceneNoCol,coltext=white,colback=sceneNoCol}
\newtcbox{\sceneNoL}{nobeforeafter,tcbox raise base, %inline
boxrule=0mm,top=0mm,bottom=0mm,right=0.15mm,left=0.15mm,arc=1.5mm,boxsep=1mm,
fontupper=\sffamily\small,
colframe=sceneNoCol,coltext=white,colback=sceneNoCol}

%% file: figures.tex
\newcommand{\figureWorkshopExample}{ % used in 4
    \begin{floatingfigure}[r]{3in}
        \noindent
        \includegraphics[width=3in]{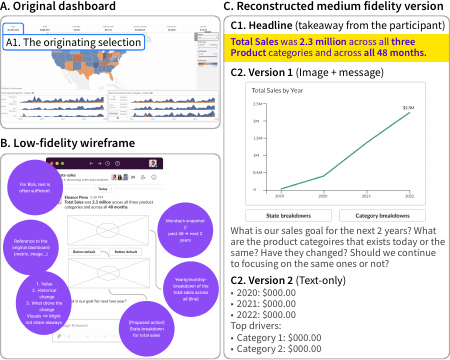}
        \caption{
        (A)~P5's dashboard for the co-design workshop. P5's data artifact originated from a selection~(A1). (B)~A low-fidelity wireframe with commentary from the session. (C)~Our reconstructed medium-fidelity version. P5 prepared the headline text~(C1) as a takeaway before the session. P5 suggested image~(C2) and text-only~(C3) designs.
        }
        \label{fig:workshop:example}
    \end{floatingfigure} 
}

\newcommand{\figureScenarioCharacters}{ % used in 4
    \begin{floatingfigure}[r]{2.35in}
        \vspace{\dimexpr0.3\baselineskip-\topskip}%
        \noindent
        \includegraphics[width=2.35in]{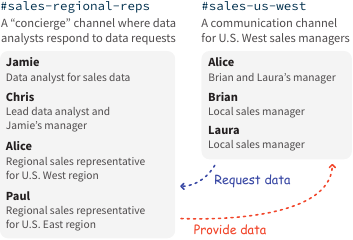}
        \caption{The characters, roles, and conversation channels appearing in our scenario.}
        \label{fig:scenario:characters}
    \end{floatingfigure} 
}

\newcommand{\figureScenarioTemplate}{ % used in 4
    \begin{figure}
        \centering
        \noindent
        \includegraphics[width=\textwidth]{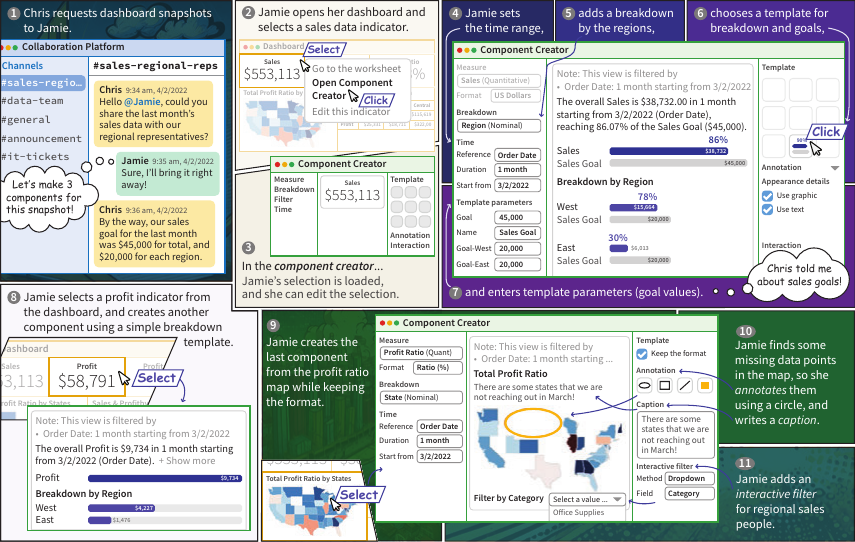}
        \caption{A scenario for creating dashboard snapshots based on our co-design workshop finding.
        }
        \label{fig:scenario:template}
    \end{figure} 
}

\newcommand{\figureScenarioComposing}{ % used in 4
    \begin{figure}
        \centering
        \noindent
        \includegraphics[width=\textwidth]{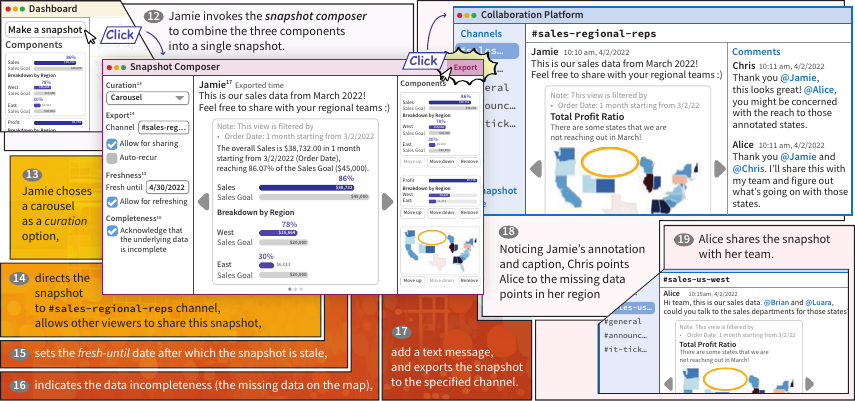}
        \caption{A scenario for composing, exporting, and disseminating a dashboard snapshot.}
        \label{fig:scenario:composing}
    \end{figure} 
}

\newcommand{\figureScenarioAutoRecur}{ % used in 4
    \begin{figure}[b]
        \centering
        \noindent
        \includegraphics[width=\textwidth]{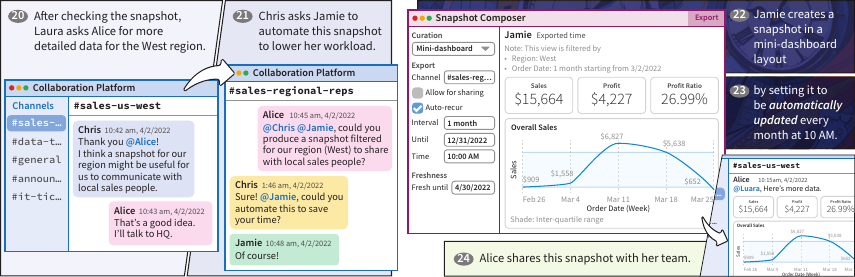}
        \caption{A scenario for setting a snapshot to be automatically updated.}
        \label{fig:scenario:autorecur}
    \end{figure} 
} 

\newcommand{\figureScenarioTelemetry}{ % used in 4
    \begin{figure}[t]
        \centering
        \noindent
        \includegraphics[width=\textwidth]{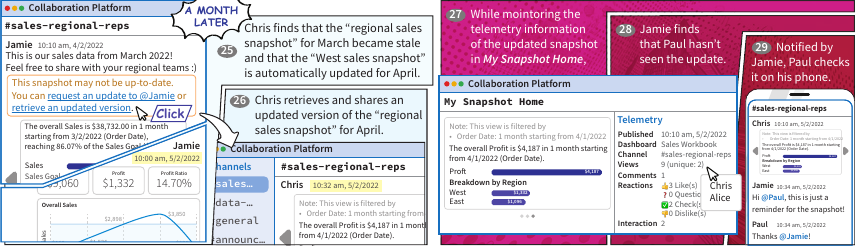}
        \caption{A scenario for retrieving updates and tracking telemetry information.}
        \label{fig:scenario:telemetry}
    \end{figure} 
} 

\newcommand{\figureFindingsOverview}{ % used in 4
    \begin{figure}[t]
        \centering
        \noindent
        \includegraphics[width=5.55in]{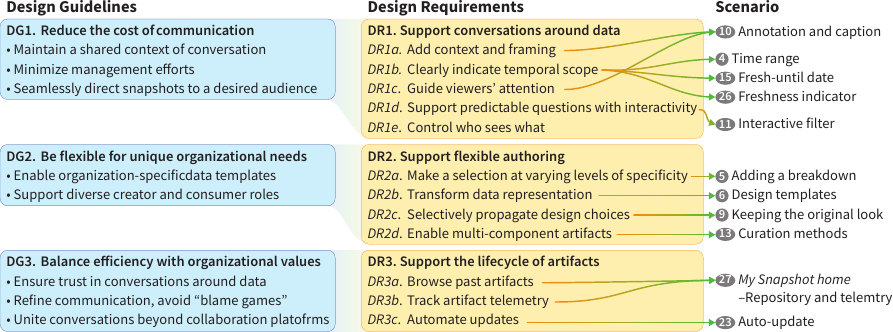}
        \caption{Our high-level design guidelines from the interview study (\autoref{sec:interview}) rationalize low-level design requirements from the co-design workshop (\autoref{sec:workshop}), manifested in the demonstration scenario (\autoref{sec:scenario}).}
        \label{fig:interview:overview}
    \end{figure} 
} 

%% file: sections/0-abstract.tex
\begin{abstract}
In enterprise organizations, data-driven decision making processes include the use of business intelligence dashboards and collaborative deliberation on communication platforms such as Slack.
However, apart from those in data analyst roles, there is shallow engagement with dashboard content due to insufficient context, poor representation choices, or a lack of access and guidance.
Data analysts often need to retarget their dashboard content for those with limited engagement, and this retargeting process often involves switching between different tools.
To inform the design of systems that streamline this work process, we conducted a co-design study with nine enterprise professionals who use dashboard content to communicate with their colleagues.
We consolidate our findings from the co-design study into a comprehensive demonstration scenario.
Using this scenario as a design probe, we interviewed 14 data workers to further develop our design recommendations. % 135 words
\end{abstract}


\ccsdesc[500]{Human-centered computing~Visualization systems and tools}

\keywords{Dashboard snapshots, visualization retargeting, collaboration platforms}

\received{}
\received[revised]{}
\received[accepted]{}

%% file: sections/1-introduction.tex
%===========================================================
\section{Introduction}
\label{sec:intro}
%===========================================================

\noindent Enterprise organizations are increasingly encouraging their employees to adopt data-driven approaches to communication and decision making~\cite{dimara2021unmet}.
This shift is affecting multiple groups across these organizations, from data scientists to managers, communicators, and other groups~\cite{brehmer2022,zhang2020data,crisan2021,tory2021datavoice,kandogan2014}.
Recently, this shift has been accompanied by a workplace culture shift to distributed and asynchronous modes of working.
To communicate across groups spanning multiple time zones and locations, collaboration platforms like Slack and Microsoft Teams have been playing increasingly important roles within organizations~\cite{zhang2020data}, gaining millions of active users~\cite{slackStat,teamsStat}.
These concurrent shifts imply a need to communicate in reference to data via collaboration platforms, particularly with those who do not occupy data analyst roles.

Posting a link to a business intelligence (BI) dashboard on a collaboration platform is unlikely to lead to fruitful conversation, as many viewers have difficulty accessing or orienting themselves within a complex dashboard or BI application.
Other common yet sub-optimal strategies include reconstructing a simpler dashboard or sharing screenshots of a dashboard or a chart.
Reconstructing a dashboard can impose extra workload, not only on dashboard authors, but also on those inclined to share insights from dashboards that they themselves did not create. 
Meanwhile, static screenshots are limited in three major ways.
First, they can become stale at any time as updated data becomes available.
Second, they may not be responsive to viewers' device or screen contexts, especially for those using a collaboration platform on a smartphone.
Finally, it can be difficult to track and update these screenshots after they proliferate across collaboration platform channels.

To empower BI application users with an effective means of communicating dashboard content to their colleagues, it is necessary to envision how to \textit{retarget} this content.
Visualization retargeting commonly refers to transforming a chart to accommodate different contexts, including audiences~\cite{Bottinger2020,Johnson2014}, device and screen types~\cite{hoffswell2020,horak2021responsive,kim:responsive2021,kim:insight2021,kim:cicero2022,Wu2013,wu2020mobilevisfixer}, and style guides~\cite{Harper2014,harper2017}.
However, retargeting dashboard content for communicative forums imposes additional challenges beyond a design space for transforming individual charts.
For example, data analysts can make a selection from the dashboard and apply a different visual format so as to more directly communicate a takeaway message. 
Furthermore, the data for a shared chart might change, requiring appropriate alerting and annotation techniques to notify both chart viewers and authors.
To explore such considerations, we conducted a co-design workshop with nine expert users of BI applications to derive \textbf{low-level design requirements}.
Given these requirements, we produce a \textbf{demonstration scenario} where a data analyst creates \textit{dashboard snapshots} to communicate with their non-analyst colleagues.
Using the demonstration scenario as a design probe, we interviewed 14 expert BI application users.
Reflecting on our findings, we contribute \textbf{high-level design guidelines} to inform the design of functionality that reduces the gap between BI applications and collaboration platforms.

%% file: sections/2-background.tex
%===========================================================
\section{Background and Related Work}
\label{sec:rw}
%===========================================================

\noindent Our work is grounded in prior work with respect to conversations around data in enterprise contexts, visualization retargeting, and collaborative visualization.

%-----------------------------------------------------------
\subsection{Enterprise Communication With and Around Data}
\label{sec:rw:snapshots}
%-----------------------------------------------------------

\noindent Across enterprise organizations, people communicate ``with and around data''~\cite{tory2021datavoice} to support conversations and inform stakeholder decisions~\cite{crisan2021,zhang2020data}.
Prior work identifies a spectrum of roles in this context~\cite{kandogan2014,kandel2012,zhang2020data}: those with high data literacy and skills for producing data artifacts (\eg~dashboards, reports) and those who mainly consume said artifacts.
We use the term \textit{data professionals} to describe the former group.
This group includes roles profiled in prior work~\cite{crisan2021,kandogan2014}, such as \textit{data engineer} or \textit{hacker}, implying professional data analysis skills like statistical modeling and visualization.
Meanwhile, \textit{consumers} often occupy management or communication-based roles and exhibit shallow and brief engagement with data~\cite{kandogan2014,zhang2020data}, prompting them to ask follow-up questions and initiate discussion with data professionals and fellow consumers. 

Data professionals and consumers make conversations and decisions that hinge on valid and up-to-date data.
Data professionals need to share data artifacts frequently and routinely~\cite{tory2021datavoice}, following updates to data, changes to analytic goals, and requests from consumers~\cite{kandel2012}.
They commonly use business intelligence (BI) dashboards to have updates with new data and adapted them in response to new goals and requests.
Dashboards created in BI applications like Tableau or Power BI often contain interactive elements, allowing users to answer specific questions by making selections using parameters or filters~\cite{sarikaya2019}.
With these capabilities, dashboards have become a popular medium for sharing observations~\cite{donoho2017}, particularly for those with limited programming expertise~\cite{kandel2011}.

Collaboration platforms, such as Slack~\cite{slackStat} and Teams~\cite{teamsStat}, increasingly support data conversation as well as daily communication within enterprise organizations.
Data professionals share links to dashboards and screenshots of dashboard content~\cite{tory2021datavoice}.
In addition, people share a high volume of links, file attachments, and images on these platforms~\cite{wang2022groupchat,handel2002groupchat}, generating a long and multimodal communication history that is tedious to search or aggregate~\cite{slack2021}.
People access collaboration platforms on smartphones, as full-screen desktop applications, or as peripheral applications monitored in the background, underscoring a need for the responsive design of shared content.

%-----------------------------------------------------------
\subsection{Visualization Retargeting}
\label{sec:rw:retargeting}
%-----------------------------------------------------------

\noindent To improve an experience of sharing dashboard content with consumers on a collaboration platform, beyond sharing links or screenshots of dashboard, we must consider ways to modify or \textit{retarget} dashboard content.
In visualization research, retargeting refers to transforming a visualization artifact initially designed for one context so that it may suit another.
This contextual shift may be due to different audience and communication goals (\eg~from a research audience to the general public~\cite{Bottinger2020,Johnson2014}) or attributed to different device sizes (\eg~responsive visualization~\cite{hoffswell2020,horak2021responsive,kim:responsive2021,kim:insight2021,kim:cicero2022,Wu2013,wu2020mobilevisfixer,kim:dupo2024}).
In case of retargeting dashboard content for reporting and sharing with consumers, data professionals commonly need to change the size and layout, compose elements with a coherent narrative, and add annotations to provide context~\cite{zhang2022understanding}.

\bpstart{Responsive visualization} 
BI dashboards are often designed for full-screen applications whereas collaboration platforms offer smaller space assets; we can apply learnings from prior work in responsive visualization design when retargeting dashboard content for collaboration platforms.
Bach~\ea~\cite{bach:2022dashboardDesignPattern} identify a dashboard design trade-off whereupon decreased screen space needs simpler representations, along with additional interactivity for revealing hidden information, in line with prior work on responsive visualization~\cite{horak2021responsive,kim:insight2021,kim:responsive2021}.
We also note that while some BI applications~\cite{tableauMobile,powerBImobile} provide a mobile app emulator to preview how a dashboard will look on a mobile display, it is difficult to simulate how dashboard content may manifest in an \textit{embedded context}, such as within a collaboration platform.

Responsive design techniques typically assume a continuity of audience, intent, and context across devices~\cite{kim:responsive2021,hoffswell2020}; for example, desktop and mobile news readers have a similar intent.
On the other hand, the intent and context may vary from one audience to another in an enterprise setting.
Data professionals need to interactively break down and (re-)interpret data content~\cite{dimara2021unmet}, while consumers tend to be more passive with shared material~\cite{tory2021datavoice,kandogan2014}.
However, this passivity should not be seen as a license to share static content; while the audience may not interact directly with the content, they may discuss it and pass it on to secondary audiences, so it is critical to keep the content updated and monitor its propagation throughout an organization~\cite{tory2021datavoice}.
Our work addresses these concerns by investigating the use of dashboard content beyond design transformation, considering also the dissemination of this content within an organization.

\bpstart{Tools for transforming dashboard content}
Prior work has investigated several aspects of transforming dashboard content.
For instance, the SoftLearn Activity Reporter~\cite{ramos2017datatotext} supports creating template-based automated text reports from dashboard content, while Codas~\cite{codas2022} supports composing report documents that integrate dashboard content with text and other media. 
Beyond reports, StoryFacets~\cite{park2022storyfacets} supports producing a slideshow from a selection of charts, and Chart Constellations~\cite{xu2018chart} can generate a summary visualization that reflects and enables collaborative data analysis.
Commercial tools such as Tableau~\cite{datastories} and Power BI~\cite{powerbistory} also offer techniques for generating reports and slide presentations from a dashboard.
Yet, these existing applications lack specific affordances for transforming dashboard content for sharing on collaboration platforms.
In particular, tools for sharing of documents and slides preclude interactivity, lacking direct affordances for an audience to ask \textit{what-if} questions~\cite{kandogan2014}.
A greater degree of interactivity is expected on collaboration platforms, where people interact with shared content while simultaneously interacting with others.
Thus, our work acknowledges the need to transform dashboard content while retaining the ability for consumers to interact with the content in curated and predictable ways.

%-----------------------------------------------------------
\subsection{Collaborative Visualization}
\label{sec:rw:collaborative}
%-----------------------------------------------------------

\noindent Isenberg~\ea~\cite{isenberg2011} propose two defining aspects of collaborative visualization: time (synchronous vs. asynchronous) and place (co-located vs. distributed).
Collaboration platforms are particularly useful for distributed organizations, and can be used both synchronously and asynchronously.
Prior collaborative visualization work suggests various techniques for such contexts. 
For example, Sense.us~\cite{heer2009} provides annotation techniques, threaded conversations, and preserved configurations to help a distributed team maintain a shared frame of reference, while
CommentSpace~\cite{CommentSpace} provides tagged comments and linked evidence to help team members easily identify others' reactions.
Although these approaches aspire to bring multi-party conversations to a data analysis platform, in practice, those in non-analyst roles often face barriers in terms of platform access and orientation, hampering their participation in such conversations~\cite{tory2021datavoice}.
Reflecting this reality, our research aims to meet people where they are already having conversations: within a collaboration platform.

%% file: sections/3-co-design.tex
%===========================================================
\section{Co-design Workshop}
\label{sec:workshop}
%===========================================================

\noindent To identify design requirements for sharing dashboard content on collaboration platforms, we conducted a co-design workshop with enterprise data professionals who use and maintain business intelligence (BI) dashboards.
In this workshop, we focused on understanding the needs of practitioners and encouraged them to look beyond the capabilities of existing commercial applications.

%-----------------------------------------------------------
\subsection{Participants}
\label{sec:workshop:Participants}
%-----------------------------------------------------------

We recruited nine data professionals (2 female and 7 male) from a multinational software company and two of its subsidiary companies.
Our participants hailed from various business areas including sales, engineering, and maintenance, with job titles including marketing manager, solution engineer, and product management associate.
During the recruitment process, each participant reported sharing data and dashboard content with their peers on Slack at varying frequencies, from once a month to several times a day.

In two instances, we held a joint 60-minute workshop session with a manager and one of their direct reports, while the five other sessions were held with individual contributors and lasted approximately 45 minutes.
All of our participants made use of Slack as a collaboration platform, as well as various dashboard tools, including Tableau, Salesforce CRM Analytics, and Looker. 

%-----------------------------------------------------------
\subsection{Protocol}
\label{sec:workshop:protocol}
%-----------------------------------------------------------

\figureWorkshopExample

\noindent \textit{Before a session}, we asked each participant to prepare one or two \textit{`takeaway'} statements reflecting the state of one of their dashboards that would be representative of something they would share on Slack.
Two participants could not share their dashboard content with us as it contained sensitive customer information, so they used a representative publicly-available dashboard familiar to them. 

\textit{During a session}, each participant co-designed a low-fidelity data artifact with us to complement their takeaway statement.
They first described how they formulated their takeaway statement, the scope of the dashboard content referenced by the statement, and whether they referenced any outside context not provided in the dashboard itself.
Next, we asked participants to rapidly wireframe a low-fidelity artifact to support their takeaway statement.
We used a Figma Jamboard~\cite{figma}, in which we provided an empty Slack interface along with preliminary design assets (\eg~text blocks, buttons) so as to better situate their data artifacts as a part of a conversation.
While we encouraged them to use Jamboard features (\eg~drawing, sticky notes), we asked them to describe their desired design and its rationale when they struggled to articulate a detailed design themselves.
We assisted by adding wireframe elements and our own notes based on their comments, as shown in \autoref{fig:workshop:example}B. 

Given that our participants were not specialized in interface design, we provided materials to help them consider different design possibilities and avoid design fixation.
To introduce a data artifact as interactive, dynamic, or non-visual content beyond a conventional static chart, hyperlink, or screenshot, we showed alternative formats like mini-dashboard, animated chart, and text-based summary to participants.
We also provided glossary cards of potential interface elements.
We further instructed them to imagine beyond any current technological limitations of their tools.

After completing the wireframing activity, we asked about the envisioned use of the data artifact they created.
First, we asked about their expectations with regards to what would happen to their artifacts after sharing them on a collaboration platform, such as data updates to the source dashboard.
Next, we asked about if and how they would monitor their audience's engagement with their artifacts.
Lastly, we asked questions regarding the integration of data artifact authoring into current workflows and their present difficulties in sharing and browsing content shared on collaboration platforms.

\textit{After a session}, we reconstructed participants' wireframes as medium-fidelity designs, giving greater clarity to participants' design ideas.
\autoref{fig:workshop:example}C illustrates an example of a medium-fidelity design. \\

%-----------------------------------------------------------
\subsection{Data Analysis}
\label{sec:workshop:analysis}
%-----------------------------------------------------------

\noindent We analyzed the artifacts generated during the workshop along with transcribed conversations with participants using a hybrid thematic analysis approach~\cite{hybridThematicAnalysis,braun2006thematic}. 
Our initial themes were derived from prior work examining enterprise data communication (\eg~\cite{crisan2021,tory2021datavoice,kandel2012}): content flexibility, time management, origins of viewing and creation, and interaction between users.
We deductively analyzed our data using these initial themes and inductively identified new themes (\eg~communication goals, telemetry viewers), eventually arriving at a set of design requirements.
The first author led the transcript coding effort and wrote initial excerpts about the codes.
As we met with the participants, we (the authors) discussed and updated the codes and excerpts.
Later, we consolidated our codes and excerpts into structured narratives.

%-----------------------------------------------------------
\subsection{Results: Design Requirements}
\label{sec:workshop:guideline}
%-----------------------------------------------------------

\noindent Our analysis yielded three sets of design requirements:

\bpstart{DR1: Support conversations around data}
Data artifacts from dashboard must become part of a conversation about data; more than the mere delivery of a chart, they must also communicate insights and calls to action.

\ipstart{\textbf{DR1a}: Add context and framing}
Some participants framed their takeaway statements and content as \textit{status reports} (\eg~\textit{``We now have} [500] \textit{people in} [CITY]\textit{''} (P3)) and as \textit{calls to action}.
For example, P8 added an informal qualitative interpretation to a status report: \textit{``Our} [server health] \textit{is taking hits in the} [US-West] \textit{environment.''} with a heatmap showing different server health values (\eg~throughput, delays) of the environment over time.
In P7's takeaway statement, the call to action \textit{``Please enter your activities on time so leadership understands the impact that we’re having.''} was accompanied by a table for different activity measures by his team members with a highlight on the member whose values were missing.

\ipstart{\textbf{DR1b}: Clearly indicate temporal scope}
To avoid misleading their audience, participants wanted to constrain their data to a temporal scope.
For example, P1 wanted to exclude the data points for June 2022 (this particular workshop session took place in mid-June), so as not to give a false impression of a lower-than-typical monthly value.
Similarly, P8 constrained data points to Mondays only because his weekend data values tended to be lower than typical weekday values, and he did not want to share unrepresentative values.

\ipstart{\textbf{DR1c}: Guide viewers' attention}
Many participants used visual emphases to indicate important data points, to specify a viewing order, or to aid interpretation.
For instance, P2 drew an arrow from one component to another to specify the viewing order.
P8 added a confidence interval for the usual range of a measure, so a big deviation could \textit{``raise eyebrows.''}
P9 included a footnote to mention missing data points.
Lastly, some participants envisioned a macro- or shortcut-based interaction for guiding viewers to specific observations:
P3, who we'll call `Andy', added a \textit{``see what Andy saw''} button that reveals a view of the dashboard with a specific combination of filter settings.

\ipstart{\textbf{DR1d}: Anticipate predictable questions with limited interactivity}
Several participants wanted to curate interactions that would allow their audience to see more details and spur further action.
For example, P9's artifact design was about \textit{``scores broken down by business unit,''} so he added a simple drop-down filter for different business units.
While adding a breakdown option to select a business category, P5 said,
\textit{``if I can just have those} [\dots] \textit{follow up questions, that would help me share holistic view of what's in my mind,''} and he would talk with his teammates when deciding interaction options for public-facing data artifacts.

\ipstart{\textbf{DR1e}: Control who sees what}
A few participants indicated the importance of governance, as data is often a sensitive asset in business contexts.
Governance for dashboards is often achieved through setting permissions (\ie~who can read, write, and delete).
For example, P9 said that he could not share dashboard screenshots on a specific collaboration platform channel because the people in that channel did not share the same permission level. 
Thus, retargeting dashboard content must go beyond static screenshots to include live visualization content through which existing data governance policies can be enforced.

\bpstart{DR2: Support flexible authoring}
Data professionals should not be bound by the form of content as it appears in a source dashboard; they must be able to transform this content in a flexible way.
However, dashboard reconfiguration or redesign can be a burden on data professionals: they often \textit{``don't have the time to go in and change things} [\ldots and] \textit{may not even have the skills to be able to go in and change it''} (P6).

\ipstart{\textbf{DR2a}: Select dashboard content at varying levels of specificity}
Many participants wanted to be able to select only part(s) of their dashboards to share a focused message.
For example, P1 indicated that dashboard content irrelevant to her message seemed to confuse people who rarely or never visited the dashboard. 
On the other end of the spectrum, data artifacts on collaboration platform could reference content spanning several disparate dashboards.
This was exemplified by P5, who selected a total sales value from one dashboard and added a temporal breakdown to it, informed by another dashboard so as to provide further context (\autoref{fig:workshop:example}A1 $\rightarrow$ C1).
Given these varying levels of specificity, we use the term \textit{dashboard selection} to refer to dashboard content that authors select to include in a shared artifact.
A selection could be a single numeric value (\autoref{fig:workshop:example}A1) or a set of charts from one or more dashboards.

\ipstart{\textbf{DR2b}: Transform how data is represented}
Some messages in a data artifact benefit from transformations to how the dashboard content is represented. 
Participants often adjusted granularity of the data to simplify the content or to provide additional context for data consumers.
For example, P2 said he would include categorical breakdowns for data consumers \textit{``to know what I’m talking about,''} because they would not be likely to visit the dashboard and apply a categorical filter themselves.
P1 wanted to disaggregate a field with a ranked list of top or bottom performers (\eg~employees, products) to make it \textit{``more tangible.''}
Finally, some participants wanted to add information not found in the the original dashboard.
For instance, P4 included the progress toward her team’s event registration goal, a value absent in her dashboard.

\ipstart{\textbf{DR2c}: Selectively propagate design choices from the dashboard}
While some dashboard content is best communicated once its representation is transformed, it was just as critical in other cases to maintain aspects of their dashboard design.
For example, P1 wanted to include a custom color legend for those who \textit{``may not be as in tune with this scale.''}
Similarly, P9 said, the color scale was \textit{``very intentional because there's someone colorblind on our team, and the colors are picked so that they have different} [meanings]\textit{. So, we can't change the colors.''}

\ipstart{\textbf{DR2d}: Enable multi-component data artifacts}
Participants often designed multi-component data artifacts, arranging these components either in a layout that mirrored the original dashboard (P2), using animation to show temporal change (P4), in a vertical stack, or as an interactive carousel of components (P7). 
Participants cited their audience's expectations when choosing how to present multiple components.
For instance, P4 mentioned that she would decide whether to use animation after assessing her viewers' impressions.

\bpstart{DR3: Support the lifecycle of data artifacts}
After sharing a data artifact on collaboration platform, authors should be able to monitor engagement with it and make updates to it.

\ipstart{\textbf{DR3a}: Browse past data artifacts}
Data artifacts on a collaboration platform channel become stale over time because their data sources can be updated at any time.
To browse their past artifacts from different collaboration platform channels, participants suggested a dedicated interface.
For example, P2 suggested a side-by-side view where one side displays a reference data artifact while the other shows all of the other past artifacts.
However, some participants said they would not browse past data artifact because collaboration platforms are for \textit{``here and now''} (P6).
Participants unanimously suggested that no changes be applied to past artifacts because, for example, changes to them may prevent people from understanding their history (P8).
P1 suggested \textit{``}[an] \textit{indicator like a gray-scale cast for freshness, without limiting people from going back and looking at the} [past] \textit{data.''}

\ipstart{\textbf{DR3b}: Track data artifact telemetry}
Participants suggested several types of data artifact telemetry on collaboration platform, such as 
(unique) view counts, name of viewers, up/down votes, interaction logs, and comments.
Ultimately, they wanted to know about the spread of their data artifacts across an organization. 
P1, for instance, was particularly interested in telemetry information because her viewers made decisions depending on information she had provided, so she wanted to adjust her content based the extent of its dissemination.
P6 suggested a distinction between \textit{active} and \textit{passive} telemetry. 
Active telemetry refers to logs of specific actions that a creator wishes their viewers to achieve, whereas passive telemetry means any kinds of logs that the system automatically records.
In addition, participants suggested where to see telemetry information:
For instance, P5 suggested a \textit{`homepage'} of sorts that summarizes engagement with shared data artifacts.

\ipstart{\textbf{DR3c}: Automate updates}
Some data artifacts are intended to be repeatedly shared as a status report.
Once initially specified, several participants wanted these to be automatically re-shared.
For example, P4 imagined automated readouts for important values (\eg~top 10 customers) \textit{``on a regular basis,''}
while P8 wanted to combine dashboard content with an automated alert system that sent out messages given predefined conditions (\eg~a quantitative value tracked in a dashboard dips below a certain threshold).

%% file: sections/4-scenario.tex
%===========================================================
\section{Demonstration Scenario}
\label{sec:scenario}
%===========================================================

\noindent 
Considering the design requirements identified in the preceding section, we introduce a demonstration scenario devoted to a \textit{``dashboard snapshot,''} which we refer to as an artifact that extends the functionality of BI dashboard applications and collaboration platforms. 
The co-design workshop participants commonly noted that the data artifacts that they shared on collaboration platforms were time-sensitive: their artifacts were relevant or valid only for a certain time period. 
Cognizant of this time-sensitivity, the term dashboard snapshot reflects how a data artifact communicates a selection of content from a dashboard for a certain time period.

To realize our demonstration scenario, we built a set of functional prototype interfaces for composing and sharing snapshots.
The intent of this scenario was to serve as a design probe for our interview study focusing on the collaborative and longitudinal aspects of enterprise conversations about data (\autoref{sec:interview}).
In this scenario, Jamie, a data professional, composes \textit{dashboard snapshots} and shares them on a collaboration platform.
Figures~\ref{fig:scenario:template}--\ref{fig:scenario:telemetry} depict the scenario.
Indices (\eg~\sceneNo{1}) indicate the corresponding parts in the figures.
We provide a video of this scenario using the prototype interface in the Supplementary Material.

%-----------------------------------------------------------
\subsection{Characters}
\label{sec:scenario:characters}
%-----------------------------------------------------------

\figureScenarioCharacters

\noindent
We introduce six characters representing different roles involved in data conversations within a large retail company, as shown in \autoref{fig:scenario:characters}.
First, Jamie is a sales analyst who works on national sales data of the company.
Jamie's manager, Chris leads the company's data analytics.
Next, Alice and Paul are sales representatives for U.S. West and East regions, respectively, dealing with plans, strategies, and goals for sales in those regions.
In a collaboration platform channel, \texttt{\#sales-regional-reps}, Jamie and Chris's teams respond to data requests from regional sales representatives like Alice and Paul.
Brian and Laura, Alice's team member, are sales managers for different areas in U.S. West. 
Alice, Brian, and Laura are members of the \texttt{\#sales-us-west} channel. \\

%-----------------------------------------------------------
\subsection{Scenario}
\label{sec:scenario:scenario}
%-----------------------------------------------------------

\figureScenarioTemplate

\bpstart{A.~Flexible authoring with templates that anticipate audience needs (\autoref{fig:scenario:template})}
\sceneNo{1}~Chris asks Jamie to share a dashboard snapshot spanning last month (March 2022) with regional sales representatives for each U.S. Region.
To make the snapshot more informative, Chris reminds Jamie of monthly sales goals per region.
For this ``regional sales snapshot'', Jamie decided to include three parts: the regional sales breakdown with goals, the regional profit breakdown, and the profit ratios (profit divided by sales) per state on a map.
\sceneNo{2}~Jamie opens her dashboard and selects a dashboard element indicating the total sales value panel consisting of an index value and its name.
\sceneNo{3}~With this selection (data and format), Jamie creates an initial snapshot \textit{component} using a \textit{component creator}.
The component creator allows for converting the format of the selection and configure properties like data filters, a time range, and additional categorical dimensions.
In the component creator, the measure (sales) is already loaded.
\sceneNo{4}~Jamie sets the time range of the component to be a month starting from March 2, 2022 (\textbf{DR1b: indicating temporal scope}). 
\sceneNo{5}~Because her original dashboard selection does not specify any regions, Jamie sets a \textit{breakdown} for the regions, which repeats values for those regions (\textbf{DR2a: selection with varying levels of specificity}).
\sceneNo{6}~As Chris has communicated the sales goals for the month, Jamie chooses a \textit{template} that includes both a breakdown and goal value, which converts the original index indicator format to a bar chart consisting of two bars for the index and goal values (\textbf{DR2b: design transformation}).
This template also adds a text summary of the information.
\sceneNo{7}~She then enters the sales goal values as \textit{template parameters} because those values are not included in the original data.
\sceneNo{8}~Similarly, she selects a profit indicator from the dashboard, and creates another component for the profit measure; in this case, she chooses a simple breakdown template.

\sceneNo{9}~For her third snapshot component, Jamie selects the profit ratio map. 
She opts to maintain this map representation in the snapshot instead of selecting a template (\textbf{DR2c: propagating design choices}).
\sceneNoL{10}~Upon specifying the March 2022 time frame for this component, Jamie notices missing data points in the West region, so she adds an \textit{annotation} and a custom component \textit{caption} to emphasize this region (\textbf{DR1a: adding context} and \textbf{DR1c: guiding viewers' attention}).
\sceneNoL{11}~Recalling that regional sales team representatives often ask her about specific product categories, Jamie adds an \textbf{\textit{interactive filter}} (\textbf{DR1d}) for the category dimension.
This interactive filter allows consumers on the collaboration platform to manipulate the snapshot on their own.

\figureScenarioComposing

\bpstart{B.~Composing and exporting snapshots with status indicators (\autoref{fig:scenario:composing})}
\sceneNoL{12}~With the three snapshot components, Jamie invokes the \textit{snapshot composer} from the dashboard application.
\sceneNoL{13}~Instead of a vertical stack, Jamie chooses an interactive carousel as a \textit{curation} option (\textbf{DR2d: multi-component snapshots}).
\sceneNoL{14}~Next, she sets the \textit{export} controls, directing the snapshot at the \verb|#sales-regional-reps| channel, subscribed to by all of the regional sales representatives.
She allows viewers to share this snapshot across other channels so that the they can communicate with their own regional teams.
\sceneNoL{15}~Cognizant that the snapshot will no longer be \textit{fresh} by the end of April, she specifies the \textit{fresh-until} date; after which, the snapshot will be indicated as stale.
She also enables other people to retrieve an updated version after the \textit{fresh-until} date (\textbf{DR1b: Clearly indicate temporal scope}).
\sceneNoL{16}~Recalling the missing data points, Jamie toggles the \textit{completeness} to indicate to her viewers that the underlying data is incomplete.
\sceneNoL{17}~After adding a text message, Jamie posts the ``regional sales snapshot'' for March, 2022 on the \verb|#sales-regional-reps| channel.

\bpstart{C.~Dissemination across a collaboration platform (\autoref{fig:scenario:composing})}
\sceneNoL{18}~Noticing the snapshot's annotation and completeness indicator, Chris determines that Alice is responsible for those missing data points in the West regions, so he mentions her in a comment.
\sceneNoL{19}~Alice then shares this snapshot with her regional team (\verb|#sales-us-west|) so that they can investigate and explain.

\figureScenarioAutoRecur

\bpstart{D.~Automating routine data communication (\autoref{fig:scenario:autorecur})}
\sceneNoL{20}~After inspecting the ``regional sales snapshot,'' Laura asks Alice for more detailed data for the West region so as to communicate with local sales agents. 
Alice redirects this request to Chris and Jamie.
\sceneNoL{21}~While agreeing with Alice, Chris also wants to reduce Jamie's workload.
Chris suggests Jamie create a snapshot for the West region's sales data with automated monthly update. 
\sceneNoL{22}~Jamie creates this ``West sales snapshot'' with three indicators (sales, profit, and profit ratio) and a line chart for overall sales trend in a mini-dashboard layout. 
\sceneNoL{23}~Jamie sets this snapshot to be \textbf{automatically updated} (\textbf{DR3c}) every month at a specific day and time. 
\sceneNoL{24}~When Jamie exports this snapshot to the \verb|#sales-regional-reps| channel, Alice shares it to her team's channel (\verb|#sales-us-west|). 

\figureScenarioTelemetry

\bpstart{E.~Reflecting on the impact of a snapshot using telemetry information (\autoref{fig:scenario:telemetry})}
\sceneNoL{25}~A month later, Chris sees that the ``West sales snapshot'' is updated and shared for April.
\sceneNoL{26}~Chris also recognizes that the ``regional sales snapshot'' for March has become stale (\textbf{DR1b: indicating temporal scope}), so he retrieves and shares an updated version.
As it was not updated by the snapshot's owner (Jamie), annotations and captions are removed so as not to mislead viewers.
\sceneNoL{27}~Meanwhile, Jamie can see the telemetry of this updated snapshot via her \textit{My Snapshot Home} interface.
\textit{My Snapshot Home} shows the list of the snapshots that Jamie has created (\textbf{DR3a: browsing past snapshots}), and each snapshot is accompanied by information about which channels this snapshot was shared to, the unique view counts and viewers, comments, reaction counts, and who interacted with the snapshot (\textbf{DR3b: tracking snapshot telemetry}).
\sceneNoL{28}~She notes that Paul (a sales representative for the East region) has not yet seen the updated snapshot, so she mentions him in a comment for the snapshot.
\sceneNoL{29}~Once notified, Paul checks out the updated snapshot via the collaboration platform's mobile app.

%% file: sections/5-interview.tex
%===========================================================
\section{Interview Study}
\label{sec:interview}
%===========================================================

\noindent 
To evaluate our initial design requirements and elicit holistic perspectives regarding the lifecycle of dashboard snapshots shared, we interviewed 14 data professionals.
We began each interview with a video of the design probe scenario described in the preceding section.
Via a video-based design probe with the prototype interfaces, we could illustrate the collaborative and longitudinal nature of the snapshot lifecycle to the participants during an interview session.
Ultimately, the goal of these interviews was to derive higher-level guidelines for supporting conversations \textit{with and around} data within an enterprise setting.

%-----------------------------------------------------------
\subsection{Method}
\label{sec:interview:method}
%-----------------------------------------------------------

\subsubsection{Participants} First, we reconnected with our co-design workshop participants, with eight of nine returning (P1--3, P5--9).
As in the initial workshop, we conducted a paired interview with colleagues P6 and P7; otherwise the interviews involved a single participant. 
As we had already established context and rapport with these participants, these interviews took 45 minutes. 
We recruited additional six data professionals (S1--6; 3 female, 3 male) from enterprise, government, and non-profit organizations to diversify our participants' backgrounds.
Their job titles included data analyst, program manager, and business analyst.
These six interviews lasted an hour, as we needed to understand their work context.

\subsubsection{Procedure} 
For our returning workshop participants, we revisited their earlier artifact designs to re-acquaint them with the context of dashboard snapshots, as two months had passed between the workshop and this study.
In general, participants found the reconstructed designs in keeping with their original intentions.
Next, we showed an 8-minute video of the demonstration scenario as a design probe.
Most participants (13 out of 14) noted that the scenario in the video reflected their day-to-day cases.
We then posed a set of interview questions regarding the utility and feasibility of the interfaces demonstrated in the video with respect to their work.
% In particular, we framed our interview questions as \textit{``How could [a feature] can be useful to your task?''} to elicit participants' different perspectives.

%-----------------------------------------------------------
\subsection{Findings: Design Guidelines}
\label{sec:interview:results}
%-----------------------------------------------------------

We thematically analyzed the interview transcripts alongside our earlier design requirements, arriving at three sets of high-level design guidelines for systems that act as communicative bridges between business intelligence tools and collaboration platforms.
\autoref{fig:interview:overview} overviews the relationships between the high-level guidelines, the low-level requirements, and the design probe scenario.

\figureFindingsOverview

\subsubsection{DG1: Reduce the cost of communication}\label{sec:interview:results:cost}
Our first set of design requirements (\textbf{DR1}) encompassed \textit{supporting conversations around data} via adding contexts of data (\textbf{DR1a}), setting a time scope (\textbf{DR1b}), guiding viewers' attention (\textbf{DR1c}), adding interactivity for predictable questions (\textbf{DR1d}), and managing permissions (\textbf{DR1e}). 
Across our interviews, we now understand a key driver behind these requirements is to reduce the cost of communication. 
Our interviewees mentioned the costs of context switching, artifact management, and an undesired proliferation of content. 

\bpstart{Maintain a shared context of conversation}
Our interviewees appreciated functionality based on requirements \textbf{DR1a--c} as ways to maintain context, and they suggested further ways to reduce communication costs in this regard.
First, there is the potential to offer more guidance when sharing content from dashboard.
For example, S6 noted that exporting dashboard snapshots with clear annotations could help people avoid unnecessary meetings and tedious deliberation.
Similarly, S1 suggested sharing \textit{``small micro videos on how} [snapshot consumers] \textit{can achieve something,''} so that they do not have to bother the snapshot creator with predictable follow-up questions or guess how a snapshot should be interpreted or used.
S5 and S6 suggested pinning a data glossary table to a channel, as terminology and reporting conventions could be local to the context of the channel. 

Aside from providing additional guidance, interviewees noted that different snapshot layout options could help guide viewers' attention more implicitly.
For example, S5 found the ability to specify the layout of a snapshot can maintain a visual correspondence between the original dashboards and the more narrowly-scoped snapshots.
Likewise, S4 said that by using a mini-dashboard layout, he could \textit{``have separate [views] created for specific audiences''} given that \textit{``it takes excruciatingly long for everything [in the dashboard] to load and [people] don't know where to look.''}

\bpstart{Minimize management efforts of creators and consumers}
While interviewees believed that self-service Q\&A via simple interactions (\textbf{DR1d}) and an automated update of snapshots (\textbf{DR3c}) could save their time, they emphasized the importance of synchronization.
For example, they described a need to synchronize sharing across different conversation channels (S3 and P1) and an ability to see updates in the same place (S3 and P3), so that people can obviate the need to browse channels in search of updates and the entire organization can stay on the same page.
P2 worried that such automation may lead to a state where people are \textit{``constantly getting spammed''} with updated snapshots.
P2 also indicated further synchronization with streaming data, such that snapshots are updated based on value thresholds instead of according to a periodic schedule. 

P3 described how self-service Q\&A on a collaboration platform could save data professionals' time.
Citing the aphorism \textit{``teach them how to fish, feed them for a lifetime,''} S6 similarly noted that providing consumers with the ability to interact with data artifacts in simple and predictable ways could preserve the attention and effort of data professionals for unique or unpredictable questions.
Similarly, most interviewees believed that indicating the temporal validity and completeness of a dashboard snapshot could help them avoid answering the same questions multiple times.

\bpstart{Seamlessly direct snapshots to a desired audience}
While our scenario included setting permissions for who can update a snapshot, it did not depict explicit features for controlling who sees what, assuming that systems can inherit organizational policy protocols.
S3 also assumed that sharing permissions are \textit{``already inherited,''} implying that data governance policies should be enforced by default.
P8 elaborated that data stewardship of this nature could relieve individual creators of the need to determine permissions for every snapshot they create.
P2 further described how a system could offer an explanation as to why specific content is obfuscated to those without permission (\eg~\textit{``they're not on the server''} [so] \textit{``they can't access the underlying data''}). 
Alternatively, P6 suggested that those lacking permissions could request special or temporary permissions from the creator to view the snapshot, for example, to support inter-organizational communications.

\subsubsection{DG2: Be flexible for unique organizational needs}\label{sec:interview:results:org}
The second set of design requirements distilled from our workshop observations pertains to the flexible authoring of data artifacts (\textbf{DR2}): selecting dashboard content at varying levels of specificity (\textbf{DR2a}), making design transformation (\textbf{DR2b}) or selectively propagating aspects of the dashboard design to snapshots (\textbf{DR2c}), and assembling multi-component artifacts (\textbf{DR2d}).
While our interviewees believed that the ability to flexibly create and share snapshots is critical in everyday communication, they also stressed the importance of how snapshot authoring should reflect organizational roles and characteristics.

\bpstart{Enable organization-specific data presentation templates}
Many of our interviewees had limited expertise in `creating' dashboards, in that they often used dashboards created and managed by others within their organizations.
Data professionals can \textit{``recycle and reuse''}~(S2) well-designed templates without having to reconstruct or redesign content created by their peers.
Reusable templates can also convey authority and elicit trust, functioning as a \textit{``stamp of approval [\ldots for] communicating certain data''}~(P7).
We learned that enterprise dashboards may incorporate domain- or context-specific design choices (\eg~a color scale used within a particular line of business within an organization).
This implies that tools that allow for recycling and reusing dashboard content should provide affordances to maintain visual design consistency, such as between a snapshot and the dashboard that it is derived from (P9).

Another aspect of organizational awareness is a need to tailor the design of data artifacts to communicate across organizational roles, to those who may have shorter attention spans and lower levels of data literacy.
For instance, S5 and S6 noted that it was often necessary to carefully align templates for presentations with the expectations of executives within their organization. 
Furthermore, visualization might not always be optimal for \textit{``users who don’t know how exactly to read [\ldots] visuals''}~(S4), suggesting the need for a greater variety of text-only templates.
As a consequence, snapshot authoring tools should provide affordances for defining organization-specific templates.

\bpstart{Support diverse creator and consumer roles}
While our scenario reflected a distinction between data professional creators and their audience, organizations consist of various data-related roles beyond this dichotomy. 
For example, S1 belonged to a three-person data team that communicated with more than a thousand employees in their entire organization.
In this case, producing snapshots for each part of the organization can be tedious, compared to other interviewees whose roles involved the in-depth use and maintenance of a small number of dashboards. 
Given the breadth of their mandate, S1 called for a self-service usage model in which consumers create their own snapshots, which assumes a certain degree of data literacy. 
On the other hand, P9 said that he often calibrated with his boss when creating important data artifacts, suggesting a possible co-creation of snapshots.
In summary, snapshot-based communication should support flexible pipelines that allow for non-linear paths with respect to creating, refining, and sharing snapshots.

\subsubsection{DG3: Balance Efficiency with Organizational Values}\label{sec:interview:results:value}
Our third set of design requirements pertained to supporting the lifecycle of dashboard snapshots once they are shared (\textbf{DR3}): providing a repository for past snapshots (\textbf{DR3a}), telemetry information (\textbf{DR3b}), and the means to automatically update stale snapshots (\textbf{DR3c}).
% The ability to specify the temporal scope of a snapshot (\textbf{DR1b}) is also important to not mislead viewers.
These requirements are reflected in our interviewees' comments relating to the mechanisms for sharing, monitoring, and maintaining snapshots, mechanisms that must reflect organizational values with respect to responsibility and trust. 

\bpstart{Ensure trust in conversations around data}
Referring to an anecdote about organizational miscommunication as a result of sharing outdated data, P1 told us that when \textit{``pulling data from my snapshot to plug into a presentation, [\ldots] it would be helpful to have the ability to say: `please don't use this data after this date'.''} 
Absent this caveat, outdated data could circulate across an organization beyond the creator's control.
In addition, S3 mentioned that \textit{``the majority of [people] don't spend their time looking for missing data,''} emphasizing the necessity of status indicators. 
Meanwhile, S5 said that indications of completeness would foster inter-organizational communication: \textit{``by the nature of the financial processing, there's some people that miss their target deadlines and the data is incomplete,''} (which is not necessarily someone's fault).
S4 also noted a need for flexibility when including these caveats, such as by adding contextual remarks.
Therefore, snapshot tools should support explicitly communicating different factors that impact the validity of a snapshot, especially for viewers who may not be inclined to question it.

Reliability and trust is also critical with respect to automated updates. 
P9 pointed out that some snapshots would require verification steps before publishing an update, particularly when they have an important executive audience. 
S6 asked if on a \textit{``Friday afternoon, you're looking at data and you have to put in a narrative about the data, is there someplace that we could put that narrative that captures that to refresh and push out on a Sunday?''}
Accordingly, functionality to defer, correct, or retract automated updates could be useful to ensure trust in shared content while reducing time and efforts for managing routine data communication.

\bpstart{Refine communication; avoid ``blame games''}
In our scenario, \textit{My Snapshot Home} offers telemetry information regarding how consumers reacted to each dashboard snapshot. 
P1 and P2 suggested using this telemetry to understand what worked well and what did not, so that creators can create better snapshots next time.
For example, telemetry might give a chance for a data professional to reason about \textit{``what tools [and] training you need''} to provide consumers, particularly for those who are not tech-savvy and \textit{``[don't] understand the platform''}~(S6).
Also, providing telemetry information should not ignite \textit{``blame games''}~(S6), which in turn creates bureaucratic overhead.
On the one hand, snapshot view counts could be used to hold decision makers accountable; on the other hand, shaming or blaming those who ignore snapshots can negatively affect collective communication.
Therefore, telemetry tools and status indicators should frame the information in a way that supports achieving communication goals (\eg~viewer engagement analysis in YouTube Studio for video creators) as opposed to blaming people in the organization. 

\bpstart{Unite conversations beyond collaboration platforms}
Organizations rely on multiple computer-mediated communication tools beyond collaboration platforms, including email (S2), wikis, cloud storage, and internal websites (S1).
Unfortunately, when dashboard content is propagated to more these or similarly static channels of communication, it becomes more difficult to respond to consumers' questions or resolve their misunderstandings in a timely fashion.
For example, S4 said that \textit{``if I were to put this in a Confluence doc [wiki], I would take a screenshot and put it there, allow[ing] me to screenshot [the temporal validity indication] as well, which tells me even in the Confluence doc.''}
We must acknowledge that dashboard snapshot tools will exist within a communication ecosystem in which conversations with and around data are taking place on multiple platforms simultaneously.
Thus, the design of snapshot creation and sharing interfaces should anticipate the static nature of email and the unidirectional communication of other channels, such as by providing affordances to redirect the conversation to more inclusive and bidirectional communication platforms. 

%% file: sections/6-discussion.tex
%===========================================================
\section{Discussion}
\label{sec:discussion}
%===========================================================

Our work complements prior work on enterprise collaboration and communication around data conversation~\cite{tory2021datavoice,crisan2021,zhang2020data,kandogan2014,kandel2011}.
Overall, prior work provided us with a baseline understanding of existing roles and practices, whereas we focus squarely on asynchronous conversations around data between distributed colleagues. 
Zhang~\ea~\cite{zhang2020data} reported how collaboration platforms like Slack support asynchronous communication between those in different enterprise roles, from data engineers to executives.
In addition to focusing on data-related conversations taking place on collaboration platforms, we also assume a prescriptive stance, providing guidelines for the design of tools that foster enterprise conversations around data.

Methodologically, we used design research methods (a co-design workshop, scenario building, and a design probe-based interview) to direct participants' attention to the many facets of enterprise communication around data and their associated application requirements.
As an outcome of this research, we presented a set of guidelines for designing tools that facilitate communication around data on threaded conversation settings.
We now reflect on the implications for existing dashboard tools and collaboration platforms, outlining necessary future work to enhance conversations around data within enterprise organizations.

\subsection{Asymmetry between Data Professionals and their Audiences}
The proliferation of extensible collaboration platforms has the potential to lower barriers with respect to engaging more people in data-driven decision making processes within organizations.
In addition to static reports that data professionals generate, consumers must be provided with more affordances to actively take part in conversations around data.
This audience group has become increasingly diverse, exhibiting varying levels of literacy and expertise.
They include executives, managers, individual contributors, and representatives from partner or subsidiary organizations.
As a consequence of this diversity, data artifacts may assume different forms, from conventional reports to sequential presentations and interactive dashboard snapshots; we learned that managing and synchronizing various forms of artifact impose significant burdens on data professionals. 

The responsibility of disseminating data artifacts often falls on a few data professionals. 
As data consumer audiences diversify, data professionals need to develop skills for data communication that are distinct from data analysis techniques.
Our findings indicate that these skills go beyond visualization retargeting to include the management of hidden communication costs, such as ensuring temporal freshness and organizational trust.
Along with the aforementioned synchronization of content across artifact types, these costs contribute to an imbalance between the time constraints of data professionals and the demands of different audience groups.
To mitigate this asymmetry, software support should distribute communication costs of data professionals to business intelligence (BI) applications and communication platforms.
In particular, the integration of existing dashboard applications and collaboration platforms can support lightweight authoring of situated data artifacts, authoring that reduces the communication burden imposed on data professionals.

\subsection{Situated Authoring of Data Artifacts}
Our design probe scenario included three distinct interfaces (the Component Creator, the Snapshot Composer, and My Snapshot Home) to clearly show the relationship between our initial design requirements and specific features. 
This separation was also convenient to direct the attention of our interviewees to focus on specific functionality.
However, in practice, the functionality contained by these interfaces may could be integrated more directly within existing BI application and collaboration platform interfaces.
This could reduce the cost of context switching remarked upon by data professionals interviewed in prior work~\cite{kandel2011,tory2021datavoice}.
To this end, we outline how existing tools can adopt functionalities to better support the routine dissemination of data artifacts.

\bpstart{Inferred data status}
In our scenario, dashboard snapshots exhibited two forms of status: freshness with respect to temporal validity and completeness with respect to missing data.
Our participants unanimously indicated that such information is vital for ensuring trust in data artifacts within their organizations.
The indicators of those statuses could be reflected in the dashboards themselves, not only in downstream data artifacts.
For example, the freshness of an artifact could be derived from its originating dashboard(s) in reference to version histories or data update cycles~\cite{heer2008graphical,loorak2018changecatcher}, yet still allowing for data professionals to manually override these derived values according to their own judgment.
In a similar way, the completeness of data represented in an artifact could encompass a wider set of data quality issues, such as missing or intentionally uncollected data~\cite{nicole2021}.
Many existing dashboard tools offers options to indicate different data quality concerns that data professionals have (\eg~VisuaLint~\cite{hopkins2020visualint}).
As our participants indicated, those concerns may depend on specific contexts, so expressing these concerns without providing related context could mislead consumers. 
Thus, an extension to dashboard tools could allow for adjusting the visibility and tone of data quality issues in downstream artifacts.

\bpstart{Flexible origin of creation}
Data professionals may have different motivations for creating data artifacts. 
As our scenario shows, they could be internally motivated to share data artifacts at a regular cadence or to share insights found serendipitously while analyzing data; in other cases, they could be externally motivated by ad-hoc requests from colleagues.
Given these different motivations, data professionals would benefit from the ability to initiate the creation of a data artifacts in different applications.
For example, collaboration platforms could directly support creating data artifacts in on-demand workflows initiated by consumers, embedding the creation interfaces as platform extensions.
Authors could copy the URL of a dashboard and paste it into a draft message, which may prompt in-situ authoring options within the collaboration platform, leveraging tools such as Microsoft Teams' Graph Toolkit~\cite{TeamsGraphToolkit} or Slack's Block Kit~\cite{SlackBlockKit}.
However, these integrated workflows would require a deeper investigation of the trade-offs between the intentionally simple and casual nature of interactions common within collaboration platforms and the steps necessary to create an artifact.
Irrespective of whether an authoring workflow is integrated into a collaboration platform message composition interface, it could be further simplified and scaffolded by leveraging automated or partially automated approaches to responsive design~\cite{kim:insight2021,kim:cicero2022,wu2020mobilevisfixer,kim:dupo2024} and for combining visualization and text~\cite{kong2012:graphicalOverlays,hullman2013:contextifier,graphscape}.
To promote greater flexibility in terms of where and how data artifacts are created, future work must connect BI and communication applications through appropriate abstractions, such as by extending dashboard specifications~\cite{epperson2023declarative}.

\bpstart{Template search and recommendation}
Future work should aim to identify a comprehensive set of communicative intents for content generated from a dashboard beyond abstract data analysis tasks, such as comparing values or determining the degree of correlation between two variables.
Once identified, additional templates may need to be designed and developed to support these intents, reflecting organizational needs (DG2). 
However, with more templates comes the need to search, filter, and recommend these templates based on the context.
Considering that state-of-the-art BI tools support Q\&A-based visualization creation using natural language queries (\eg~Ask Data in Tableau~\cite{askdata} and Q\&A box in Power BI~\cite{askdataPowerBI}), future work could enable searching and recommending templates by asking authors to write a caption first. 
Alternatively, template recommendations could draw from a summarization of the conversation that has recently taken place on the collaboration platform channel~\cite{zhang2018making} where the data artifact is to be posted.

Another possibility is the prediction and modeling of visualization intents for existing dashboards and their component views. 
For instance, Pandey~\ea~\cite{pandey2022medley} describe several analytical intents of dashboards, including measure analysis, change analysis, category analysis, and distribution analysis.
If the intent of an artifact is apparent from the dashboard that it originates from (\eg~via title or annotation), this intent could be used to constrain a likely set of templates.

\bpstart{Supporting alternative modalities}
Upon interpreting our findings from our co-design workshop, we realized that design transformations made to dashboard content are not limited to transformations of visual representation, but also transformations between visual representation and text.
Data artifact authoring could be complemented by the automatic generation of data summaries, data facts, or data stories as envisioned in prior work such as Voder~\cite{Srinivasan2019voder}, Explain Data~\cite{explaindata}, and Calliope~\cite{shi2020calliope}.
Additionally, this auto-generated text could reference specific visualization elements as proposed by Mathisen~\ea~\cite{insideinsights2019} and Kong~\ea~\cite{kong2014extractText}.

\bpstart{Integrated data governance}
Granting viewing permissions to data artifacts was discussed in our co-design workshop and again in our interview study, emphasizing the need for a common data governance model across dashboard tools, collaboration platforms, and retargeting interfaces.
Data governance refers to a set of policies that secure data and grant authority over these policies, determining data access at various levels of detail and throughout the lifecycle of data~\cite{khatri:2010dataGovernance,abraham:2019dataGovernance}.
To enable data governance for an ecosystem that includes communicative data artifacts, tools within this ecosystem need to share common expressions for relevant permissions (\eg~Windows Server~\cite{windowsAccessControl} or Amazon Web Service~\cite{awsIAMpolicy}), particularly given that collaboration platforms and BI tools traffic in sensitive organizational information.

\subsection{Limitations}
While we engaged with enterprise data professionals in both our co-design workshop and interview study, we did not directly involve data consumers.
Instead, we asked our data professionals to serve as proxies for their respective audiences since they were, in most cases, audiences themselves in conversations around data initiated by other data professionals within their organization.
Thus future work could involve those in management or executive roles as well as anyone who exhibits shallow engagement with data~\cite{kandogan2014,zhang2020data}.

%===========================================================
\section{Conclusion}
\label{sec:conclusion}
%===========================================================

\noindent Conversations around data taking place within enterprise organizations are thwarted by a missing connection between data analysis tools and collaboration platforms. 
To investigate new ways by which collaboration platforms could host data conversations, we began by conducting co-design workshops with nine enterprise data professionals to identify low-level design requirements.
These initial requirements encompassed the integration of data artifacts into conversations, the ease and flexibility of authoring data artifacts, and the management of the data artifact lifecycle.
With these requirements in mind, we produced a demonstration scenario depicting the enterprise lifecycle of \textit{dashboard snapshots}, from their authoring to their dissemination and monitoring on a threaded collaboration platform.
Leveraging this scenario as a design probe, we interviewed co-design workshop participants along with six additional data professionals.
Our interview findings suggest a greater need for managing communication costs between data professionals and consumers, for flexible data artifact templates that accommodate the diversity of roles across an organization, and for balancing communication efficiency with organizational values of trust and reliability.
Finally, in reflecting on our findings, we discussed the implications for existing dashboard applications and collaboration platforms.